\documentstyle{cupconf}
\input psfig

\ifoldfss
\else
  \ifnfssone
    \newmathalphabet{\mathit}
      \addtoversion{normal}{\mathit}{cmr}{m}{it}
      \addtoversion{bold}{\mathit}{cmr}{bx}{it}
    \newmathalphabet{\mathcal}
      \addtoversion{normal}{\mathcal}{cmsy}{m}{n}
    \else
    \ifnfsstwo
    \fi
  \fi
\fi

\def\hexnumber#1{\ifcase#1 0\or1\or2\or3\or4\or5\or6\or7\or8\or9\or
 A\or B\or C\or D\or E\or F\fi }

\makeatletter
\ifx\CUP@mtlplain@loaded\undefined
\else
\fi
\makeatother
 \makeatletter
 \ifx\CUP@mtlplain@loaded\undefined
   \font\tenbmi=cmmib10 at 10pt
   \font\sevenbmi=cmmib10 at 7pt
   \font\fivebmi=cmmib10 at 5pt

   \newfam\bmifam
   \textfont\bmifam=\tenbmi
   \scriptfont\bmifam=\sevenbmi
   \scriptscriptfont\bmifam=\fivebmi
   
 \fi
 \makeatother
\ifnfsstwo

\fi
\ifnfssone

\fi
\ifoldfss

\fi

\mathchardef\varLambda="0103

\makeatletter
\ifx\CUP@mtlplain@loaded\undefined
\else
\fi
\makeatother
\makeatletter
\ifx\CUP@mtlplain@loaded\undefined
  \font\tenbms=cmbsy10
  \font\sevenbms=cmbsy10 at 7pt
  \font\fivebms=cmbsy10 at 5pt
  \newfam\bmsfam
  \textfont\bmsfam=\tenbms
  \scriptfont\bmsfam=\sevenbms
  \scriptscriptfont\bmsfam=\fivebms

  \edef\bsy@{\hexnumber\bmsfam}
  \mathchardef\bnabla="0\bsy@72
\fi
\makeatother
\title[HDF: Introduction \& Motivation]{The Hubble Deep Field: Introduction and Motivation}

\author[Richard Ellis]%
{R\ls I\ls C\ls H\ls A\ls R\ls D\ns S.\ns E\ls L\ls L\ls I\ls S$^1$}

\affiliation{$^1$Institute of Astronomy, Madingley Road, Cambridge
CB3 0HA, England. email:rse@ast.cam.ac.uk}

\begin{document}
\ifnfssone
\else
  \ifnfsstwo
  \else
    \ifoldfss
      \let\mathcal\cal
      \let\mathrm\rm
      \let\mathsf\sf
    \fi
  \fi
\fi

\maketitle

\begin{abstract}
Although it is just over a year since the data was made public, the HDF
exposure has stimulated considerable progress towards our understanding
of the faint galaxy population. I present a brief personal account of
the history of faint galaxy studies culminating in the HDF, and describe
what I consider to be the main highlights thus far from this remarkable
image. The HDF has given considerable impetus to studies of galaxy
evolution and this has led to the emergence of a convincing empirical framework.
Further exploitation of deep HST images in conjunction with ground-based 
2-D spectroscopy will assist in the physical understanding of the 
evolutionary processes involved.
\end{abstract}
   
\section{INTRODUCTION}

We're here to celebrate and discuss scientific results from the
Hubble Deep Field (Williams et al 1997). Most would agree that this 
exposure represents an observational landmark in the long arduous path 
of exploring and understanding the Universe of faint galaxies. Indeed, 
it is difficult to remember a single observation in astronomy that 
has influenced our subject so quickly. Moreover, its full impact may 
not yet be realised. In this workshop, we will debate the significance 
of the conclusions so far derived and learn of new developments that 
follow directly from the HDF. This remarkable image has acted as an 
inspiration to many astronomers because, exceptionally, we were granted
immediate access to the data. Those working with other facilities, 
such as ISO (Rowan-Robinson, this volume) and the VLA (Kellermann, 
this volume) have been quick to follow the example by concentrating 
their deepest exposures on this same field.
 
As well as explaining what I consider to be the main extragalactic 
highlights from the HDF at this point, largely to set the scene for the more 
detailed articles that follow, I will recall some of the earlier work 
which inserted pieces of the jigsaw that we now recognise more clearly 
via the HDF. Of course, HDF has also had significant impact in the
non-extragalactic area. I'm glad this is well covered in the workshop
but won't attempt to review progress in those important areas.

\section{HOW THE HDF CAME TO BE}

How did the HDF come to be and why has it been so successful? Bob
Williams convened an advisory committee comprising a number of
extragalactic scientists who met here on 31st March 1995. Our brief was
to be visionary and to consider the optimum science that would emerge
from a significant allocation of Director's Discretionary Time. Those
present will remember a rather rambling discussion which focused
ultimately on one or two long exposures, both as a scientific mission
and a public legacy for HST. My own notes from that meeting indicate
much disagreement on details: the number of fields (north and south or
just one?), the number of filters (surprisingly only one or two..nobody
argued for more so far as I can recall), where to point (interesting
area with a distant QSO or cluster, or a blank field?). Some of us
questioned whether the community should at least be allowed to
demonstrate whether it had a smarter idea than those of the gathered
`experts'. The latter proposition, unsurprisingly, did not achieve much
support! What I am trying to say, in a way that does not insult my
fellow committee members, is that we hardly prescribed HDF at that
meeting. The credit lies with Bob Williams and his team at STScI who
turned a very sketchy idea into a carefully-planned series of
observations.

Contrary to what some might imagine, the reason HDF has been so
successful is not solely the depth of the image; images almost as deep
have been obtained from the ground (c.f. Metcalfe et al 1995, Smail et al 
1995). The major step forward was the combination of the image quality 
only HST can deliver and, foremost, the multi-passband nature of the data. 
The use of 4 strategic filters, whose relative exposure times were 
carefully balanced, has been particularly successful. The UV and blue exposure 
times were prohibitively long for most guest observers and surprisingly 
little multicolour data had been obtained with HST prior to the HDF. Add 
to that the public interest in the beautiful colour images, free access 
to the data, the galvanising effect of this large investment on 
astronomy's premier facility on other telescopes and you have the 
ingredients of the success of the HDF.
 
\section{THE HIGHLIGHTS}

So what are the most important scientific results from HDF so far? I
only have sufficient time to sketch what I consider to be the five most
significant results whilst giving due credit to earlier workers
who established the foundations of what has emerged. I have to be a bit
selective so I present this as a personal account rather than a 
comprehensive review.
 
\subsection{The flattening of the count slope N(m)}

The quest to take deeper images of the sky motivated the HDF at its
most basic level but this quest has a long and distinguished history.
Sandage (1995) discusses the classical work, whereas Koo \& Kron (1992)
and Ellis (1997) review the more recent observations. The modern era 
begins with the commissioning of the wide-field prime focii on our 
national 4-m telescopes in the 1970's. Combining fine-grain emulsions 
and automatic measuring machines, Kron, Kibblewhite and Tyson laid the foundations of image processing of faint galaxy images (Kron 1978, Peterson
et al 1979, Tyson \& Jarvis 1979). In an era when the photographic plate 
is so often disregarded, it is salutory to note how much of our 
observational achievement was established from photographic plates. Only 
now, after 20 years, are giant CCD cameras rivalling the combined depth 
and field of view. Of particular note for this meeting is Koo's thesis 
(1981) where, in an early version of the HDF, four-colour photographic
photometry was analysed in the context of photometric redshifts to 
demonstrate enhanced star formation as a function of look-back time.
 
\medskip
\centerline{\psfig{file=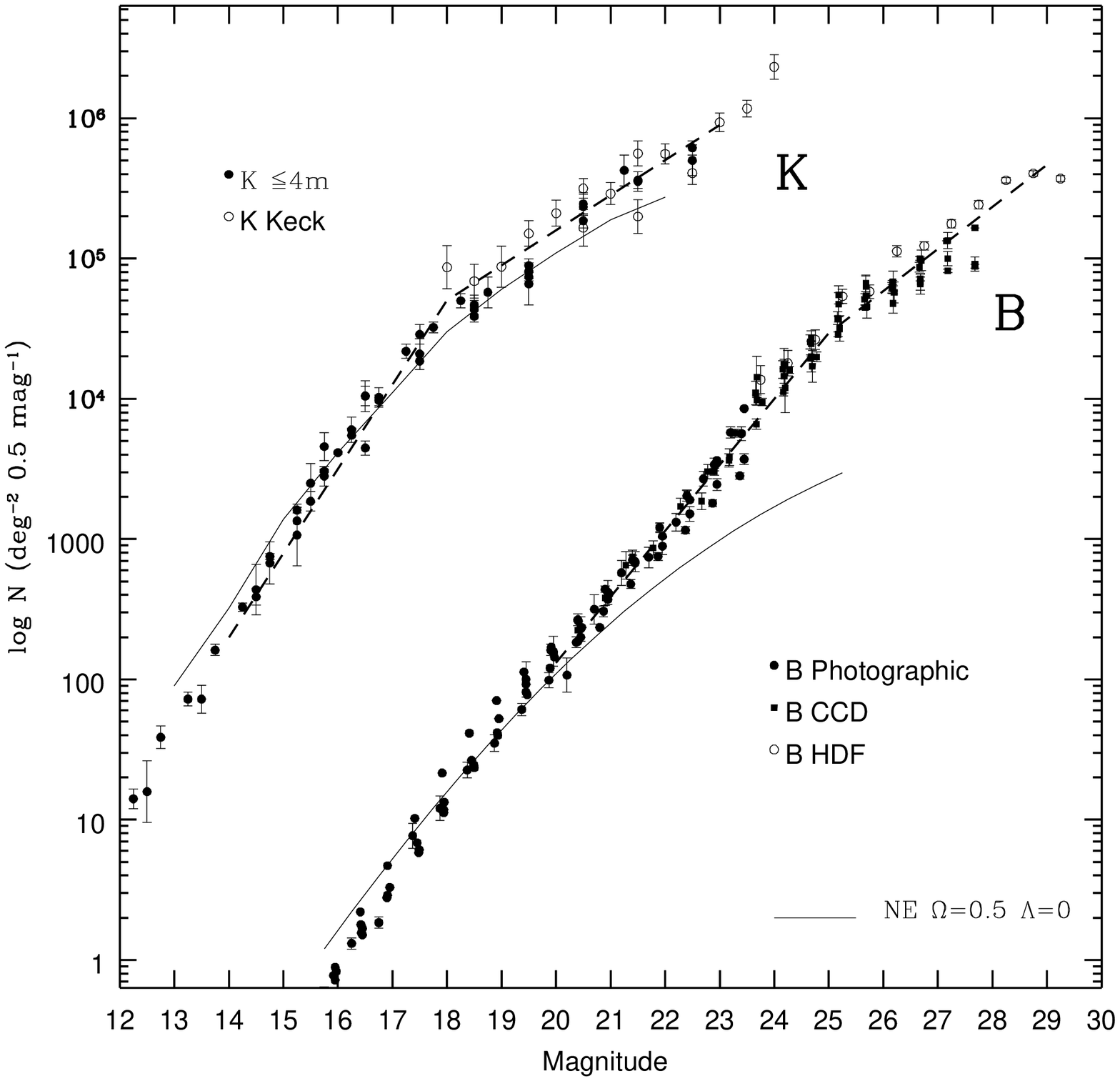,width=11cm}}
\noindent{{\it Figure 1: Differential number magnitude counts in the
B-band (including those derived from the HDF) and K-band (including
recent Keck determinations using the Keck telescope). The K counts have
been offset by +1 dex for clarity. The two power law slopes (dashed
lines) indicate the point beyond which the integrated night sky
brightness begins to converge. Solid lines refer to no evolution
predictions in the Einstein-de Sitter case (see Ellis 1997 for further
details).}
\medskip

The importance of pushing deeper was obviously recognised. Tyson (1988)
was the first to attempt ambitious long CCD exposures developing,
with Jarvis, Valdes, Seitzer and others, the relevant observing and processing
technologies. The steep blue count slope first found by Kron (1978) 
seemed to continue and many of us imagined we might soon hit the 
confusion limit. The suggestion that the count slope flattened below 
the Olber's limit, $dlogN/dm$=0.4, came tentatively from Lilly et al 
(1991) and later, with greater confidence, from the very deep exposures 
taken by Metcalfe et al (1995) including the {\it Herschel Deep Field}, 
a friendly ground-based rival of the HDF\footnote{I 
should add, somewhat topically, that Tom Shanks looked like doing 
marvellously well out of the recent British election as his Herschel Deep 
Field blue galaxies were adopted as a slogan by the Conservative 
party whose emblem was projected on them in a national newspaper. However,
the colour of the British sky has since switched dramatically from blue to 
red!}.

The flattening was dramatically confirmed in the HDF counts and
overcounting multi-component galaxies as separate units (Colley et al
1996) would make the faint slopes even flatter. The bulk of the received
extragalactic light must therefore come from the point of inflexion -
an apparent magnitude ($B\simeq$25) within spectroscopic reach where the mean
redshift is modest (z$\simeq$1). Importantly, the same effect has
been seen in the infrared at $K\simeq$18 (Gardner et al 1993, Moustakas 
et al 1997). Very few of the faint K-selected galaxies are not visible 
in the optical suggesting the result is a fundamental feature of 
galactic history.

\subsection{The small angular sizes of the faintest galaxies}

On first seeing the HDF image (in a national newspaper) I was struck 
by the amount of blank sky it contained. Together with the flat count slope, 
an  important secondary conclusion arising from this simple
observation is the small angular sizes of the faintest sources.

\medskip
\centerline{\psfig{file=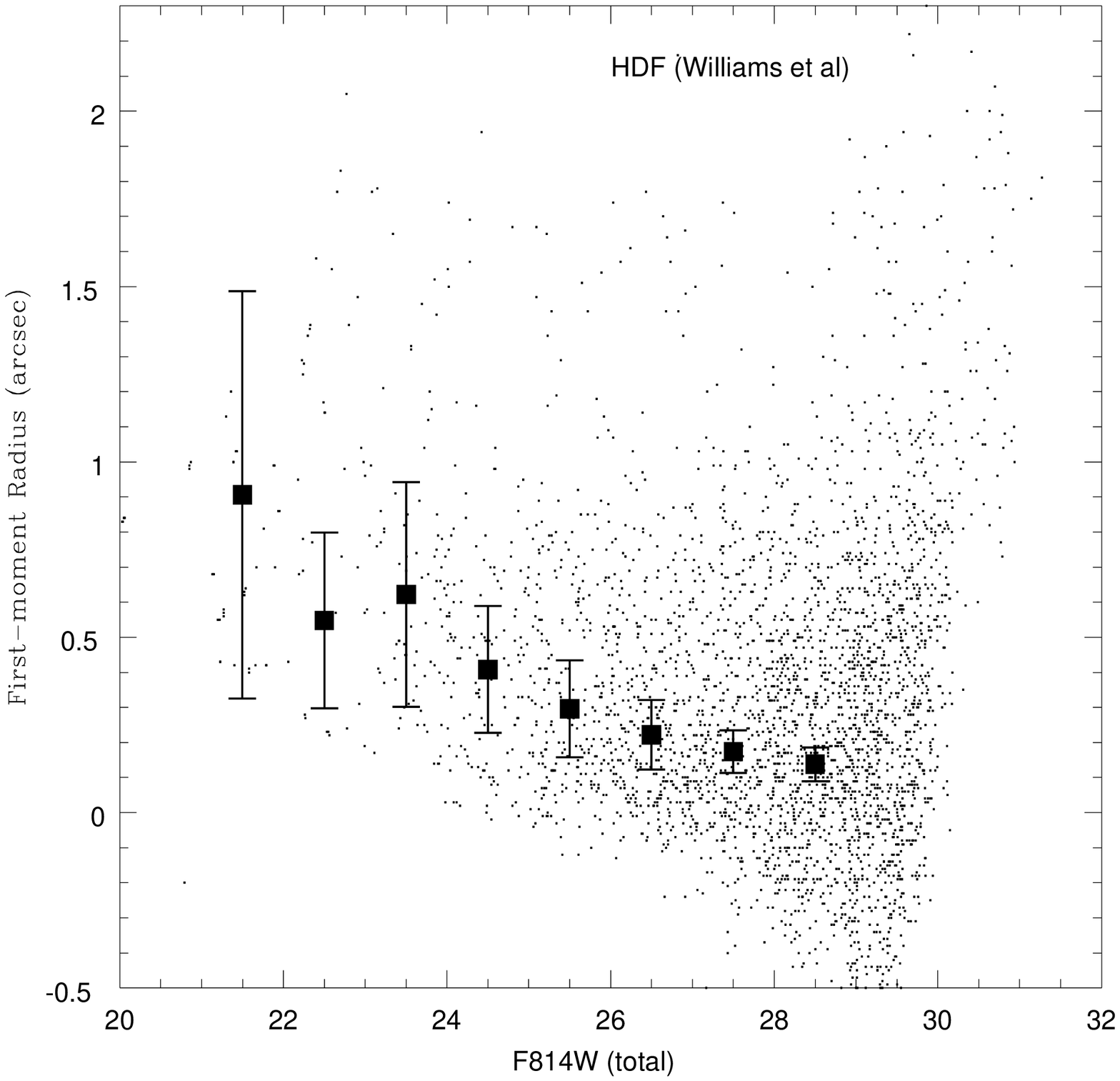,width=9cm}}
\noindent{{\it Figure 2: Correct intensity-weighted first moment 
radius versus apparent magnitude with population median and standard 
deviations for sources in the HDF.}
\medskip

This is a newer result mainly due to HST but one whose interpretion is
perhaps less straightforward (see Ferguson, this volume). Most of the
early ground-based photographic and CCD data was taken in what is now
considered to be mediocre seeing. Astronomers at the NTT and CFHT showed the
importance of improving the dome seeing and demonstrated that most of the
accessible faint blue galaxies brighter than the count slope `break point'
($B<$25) were resolved (Giraud 1992, Colless et al 1994). The first deep 
Keck images (Smail et al 1995) and HST Medium Deep Survey images (Roche 
et al 1996) suggested half-light radii of $<$0.3 arcsec at fainter limits. 
HDF has extended this trend to considerably fainter limits (Fig.~2). 

The result has a simple interpretation in the context of hierarchical
galaxy formation since, beyond $z$=1, a small angular size corresponds
to a physically small source ($\simeq$2-3 $h^{-1}$\,kpc) regardless of the
cosmological model. On the other hand, surface brightness losses and
effects due to band shifting may conspire to reduce extended sources
to apparently point-like HII regions at high redshift. Coaddition of
representative cases suggests this is unlikely to be the case. However,
NICMOS images may give a more representative indication of the physical
sizes involved.

\subsection{The increasing fraction of irregular and multiple component systems}

The 1980s also saw the first deep redshift surveys which provided
quantitative evidence for galaxy evolution brighter than the break
points in Fig~1. Progress followed the new technology of multi-object
spectrographs, from plug-plate fibre systems (Hill et al 1980), robot
positioners (Parry \& Sharples 1988) to multislit spectrographs such as
the Cryogenic Camera (Butcher 1982), LDSS-1/2 (Colless et al 1992,
Allington-Smith et al 1994) and MOSIS (LeFevre et al 1994). Such
surveys revealed a rapidly declining population of star-forming
galaxies over 0$<z<$1 which seem to be responsible for the bright count
excess. What are these rapidly-evolving galaxies?

The Medium Deep Survey (Griffiths et al 1994, Windhorst et al 1995)
analysed parallel WFPC-2 data for $\simeq$30 fields and presented the
morphological mixture as a function of apparent magnitude. Although
normal galaxies are seen in numbers consistent with approximately
constant co-moving space densities, these authors were struck by the
rising fraction of faint system with irregular morphology; many are
suggestive of merging systems. Glazebrook et al (1995b) and Driver et al
(1995) argued that rapid evolution was almost exclusively occurring in the
`irregular/peculiar/merger' category -- admittedly rather a catch-all for
non-regular systems whose physical nature remains unclear. Abraham
et al (1996a,b) introduced a more quantitative basis for faint galaxy
morphologies and discussed how to allow for redshift-dependent distortions
in extended the analysis, using the HDF, to I=25. They claimed few of the
faintest galaxies could be shoe-horned into the classical Hubble sequence.

\medskip
\centerline{\psfig{file=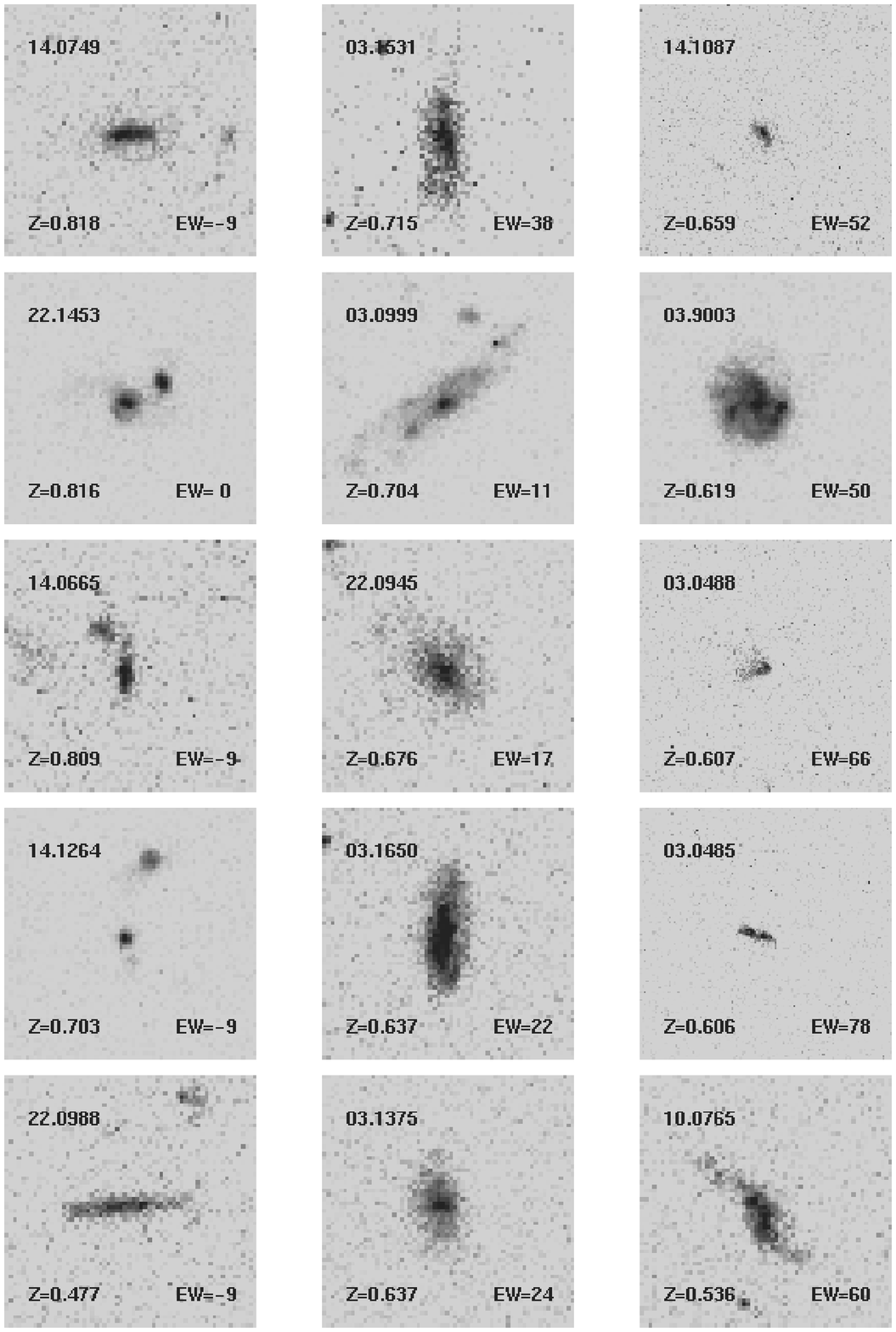,width=9cm}}
\noindent{{\it Figure 3: A mosaic of HST images for galaxies with 
irregular morphology selected from the sample of Brinchmann et al
(1997). Labels refer to rest-frame [O II] equivalent widths and redshifts.}
\medskip

The linking of redshifts and HST morphologies has been slow to emerge.
An international group, based on the original CFRS and LDSS teams, have
now amassed over 300 faint galaxies to $I<$22 or $B<$24 for which extended
WFPC-2 imaging and spectroscopic redshifts are available. This was not
feasible to construct with the Medium Deep Survey because of the mismatch
between the small WFPC-2 field and the larger ones of the ground-based
multiobject spectrographs. Brinchmann et al (1997) examined in some
detail the possible effects of redshift on the perceived morphology.
They found a rapid rise with redshift in the fraction of galaxies with
irregular morphologies which they claim cannot be due to k-correction
or surface brightness effects. This trend represents a major component of the
evolution seen over 0$<z<$1. However, it is clear the evolution in blue
luminosity density need not come entirely from this population (Lilly,
this volume) and, moreover, it is not obvious that the declining fraction
of galaxies with irregular sources is matched by a compensating growth
of disks or spheroidal galaxies over the same period as might be expected
in simple hierarchical pictures (Baugh et al 1997).
 
\subsection{The location of high redshift galaxies with modest star 
formation rates}

The most important feature of the HDF was the dedication of a significant
fraction of the Discretionary Time to deep UV and blue exposures useful
for locating the high redshift galaxies. Using photometry to constrain
the redshift distribution from the effect of the Lyman limit was first
attempted by Guhathakurta et al (1990).  They showed that the bulk of the
$R<$25 sources most probably had redshifts $z<$3. Steidel \& Hamilton
(1992), Steidel et al (1996a) later demonstrated prior to the HDF via
their own imaging and Keck spectroscopy in QSO fields with Lyman limit
absorption line systems.  The broader significance of this work was
amplified and extended to lower redshift using the shorter wavelengths
available with the HDF. Although Giavalisco et al (1996) had the first HST
images of the z$>$2.8 galaxies, the quality of the HDF images selected
to be beyond $z$=2.3 by a similar technique was considerably superior
(Ellis 1996, Fig.~3).

\medskip
\centerline{\psfig{file=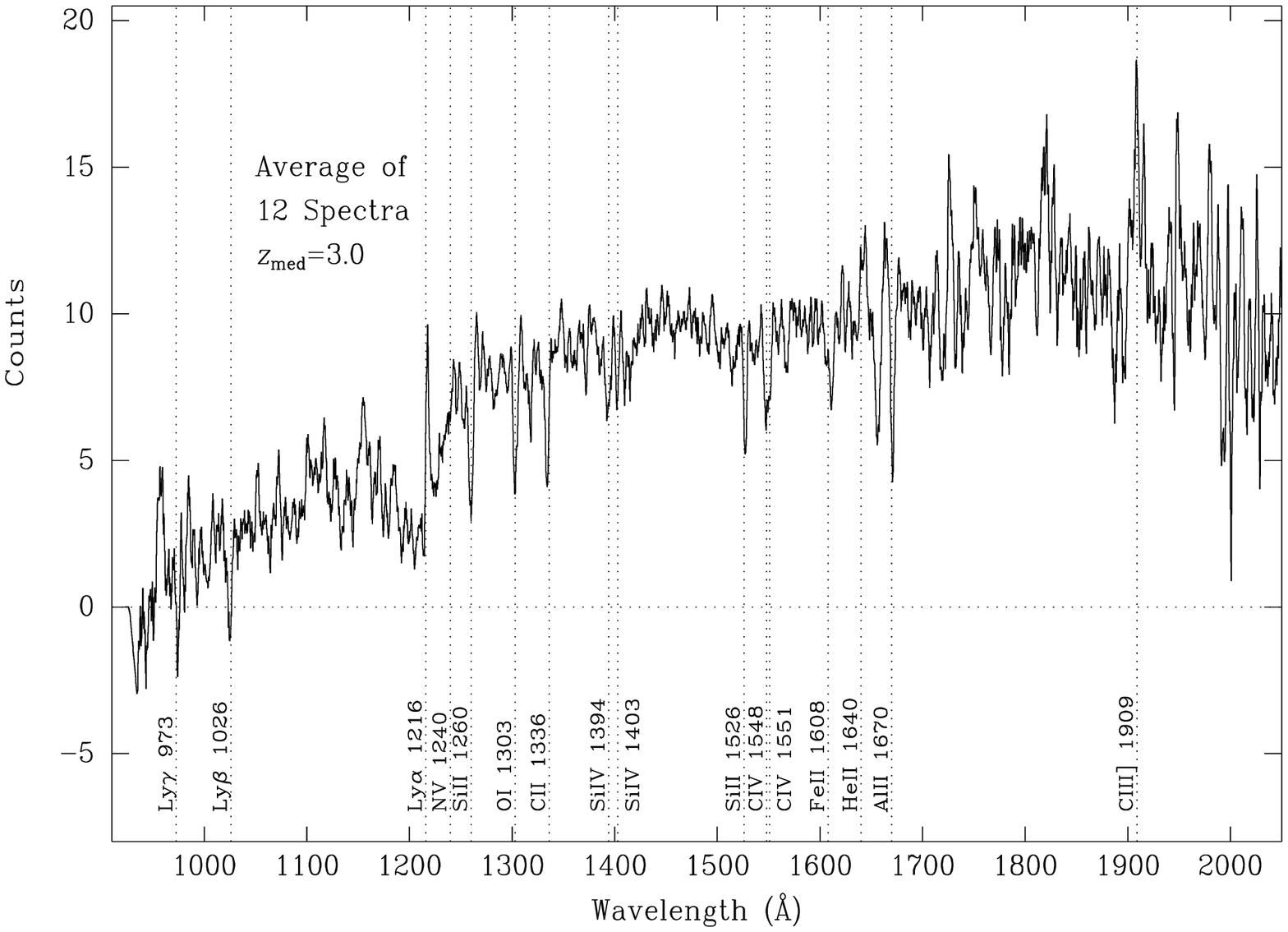,width=8cm,angle=270}}
\noindent{{\it Figure 4: Mean spectrum of the U dropouts in the HDF
and flanking fields from the analysis of Lowenthal et al (1997).}
\medskip

An interesting feature of this remarkable population is its relatively
low abundance (compared to the huge number of foreground systems) and
the modest inferred star formation rate derived from the ultraviolet
continuum flux. Other speakers will elaborate on progress in this area
and no doubt the precise star formation rates and the possibilities of
misinterpretation will be raised. Enough spectra have now been taken
(Fig.~4) for us to realise they are mostly metal poor precursors but
of what kind of galaxy and with what mass is not yet clear. The crucial
point for now is that via these and other observations, the volume density
and spectral characteristiscs of star-forming galaxies at z$>$2.3 have
become available and confirmed what was suspected from the counts,
sizes and morphologies: we are probably looking at the first era of
star formation in some category of galaxy. Most of the activity which
produced the regular Hubble sequence occurred at lower redshift.
 
\subsection{Constraints on the redshift distribution of the faintest systems}

The multicolour HDF data has led to a resurrection of interest in
estimating redshifts from colours. The technique goes back to Baum
(1962), Koo (1985) and Loh \& Spillar (1986). It is interesting to
note that the method of photometric redshifts was heavily criticised
by some at that time even though, in the case of Loh \& Spillar, more
than 4 filters were employed\footnote{I remember at a conference in
Erice there were separate dining arrangements for those who believed
in photometric redshifts!}.  However, I think some people missed
the point. In my opinion, the motivation is not to predict a precise
redshift as a substitute for a spectrum, but rather to use the method
to secure an overall statistical distribution $N(z)$. Unfortunately 
the predominantly blue SEDs make this difficult to achieve using optical 
data at 1$<z<$2 because of the paucity of spectral discontinuities. 
Comparisons amongst the various HDF photometric redshift catalogues
(Ellis 1997) shows no convincing evidence that the redshift resolution 
with 4 optical filters is any better beyond $z\sim$1 than that associated 
with the Lyman limit moving through the filters. Of course, that is 
already important information! Moreover, the addition of near-infrared 
data should improve the situation considerably (Lanzetta, this volume). 
The first results from Connolly et al (1997) are particularly interesting 
and suggest a mean redshift to $J$=23.5 that is surprisingly low, in 
rough agreement with that inferred to $R$=25.5 from the lensing inversion
technique of Ebbels et al (1997) when applied through rich well-constrained
clusters such as Abell 2218 (Kneib et al 1996) (Fig.~5).

\medskip
\centerline{\psfig{file=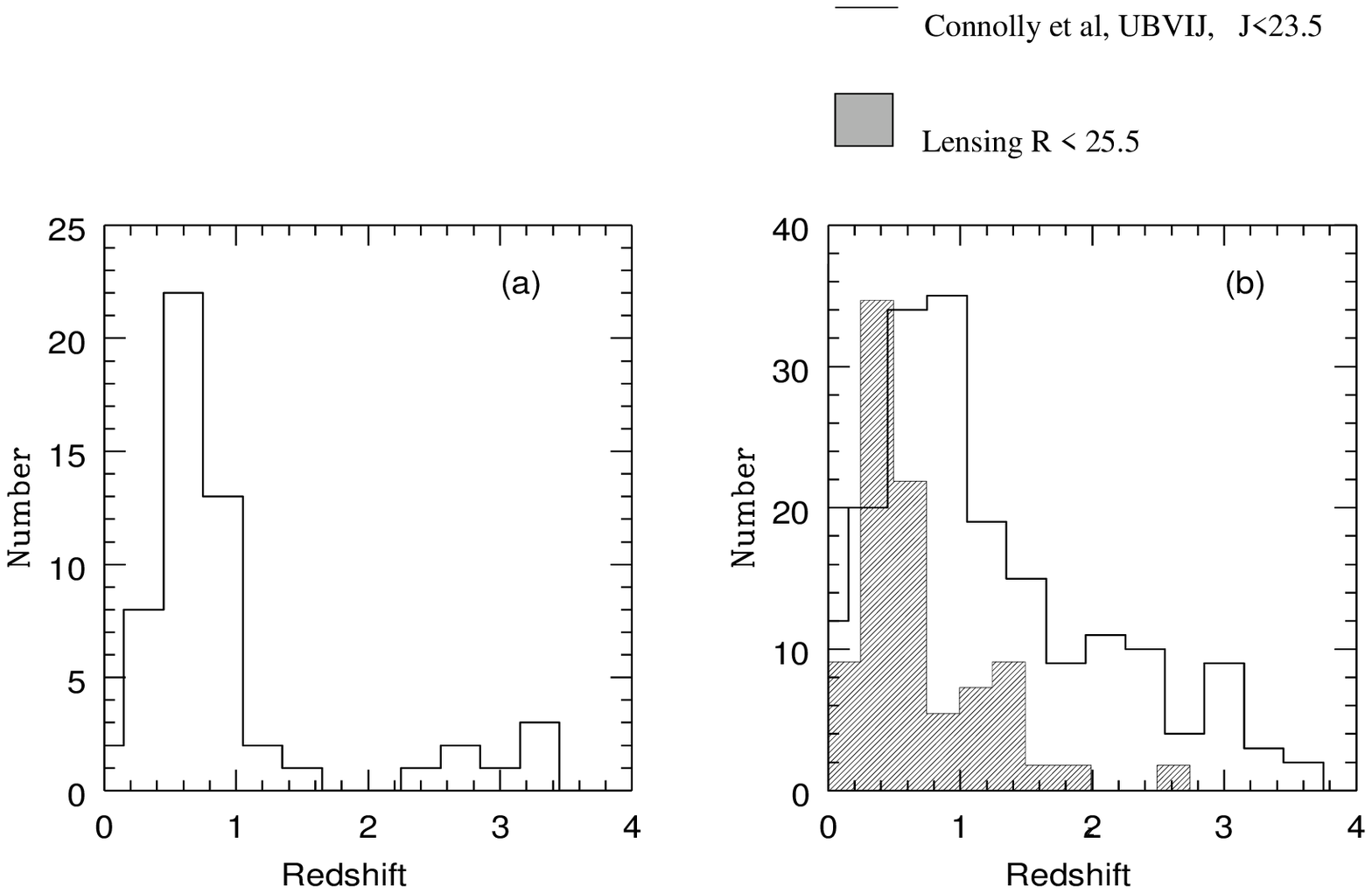,width=8cm}}
\noindent{{\it Figure 5: Redshift distributions of very faint galaxies:
(a) using Keck spectroscopy inthe HDF but to no formal magnitude limit, 
(b) photometric redshifts to $J$=23.5 utilising HDF 4-colour photometry 
and ground-based JHK photometry from the analysis of Connolly et al (1997). 
The shaded distribution is that inferred to $R$=25.5 from gravitational 
lensing viewed through the well-constrained cluster Abell 2218 (Ebbels et 
al 1997). Regardless of the technique, a surprisingly large fraction of 
the population beyond the break point in the counts (Fig.~1) has $z<$1.}
\medskip

\section{THE UPSHOT}

The above results, which of course by no means come exclusively from
HDF (but have benefitted enormously from it), point to a remarkably
simple observational synthesis discussed by Fall et al (1996), Madau et 
al (1996) and Madau (1997a,b). This will no doubt be widely discussed
during the meeting. The observational data point to a remarkably recent 
era of major star-formation as delineated by the radiation we can
see. Although different techniques are used to delineate the total star
formation rate per comoving volume and extrapolation is necessary beyond
the magnitude limits probed at the various redshifts, progress is already
being made to overcome these limitations. Supporting the argument that
we have witnessed the construction of galaxies directly with HST over
1$<z<$3 are the flattening of the faint K-band counts (which precludes the
existing of a significant population of highly reddened sources) and the
rapid growth in mean physical size, and in the 0$<z<$1 era, the declining
fraction of irregular and multi-component galaxies in conjunction with the
luminosity function changes witnessed in the redshift surveys. Completely
independent support comes from trends with redshift in the gas content
(Storrie-Lombardi et al 1996) and metallicity (Pettini et al 1994)
of the QSO absorption line population. There are many uncertainties in each
of these measures but the synthesis of so many results from different 
directions is quite compelling.
 
We should remember that this an {\it empirical} picture; it does not 
guarantee an unique physical interpretation. Much attention has
been given recently to a supposed theoretical triumph in explaining it
on the basis of hierarchical models (Baugh et al 1997). I think we have
to put this result in perspective. Perhaps it is not surprising that
hierarchical models can be arranged to approximately fit the above 
observations given the redshift of peak star formation activity is surely
sensitive to the precise way in which feedback is implemented in 
the models\footnote{Actually the fit of Baugh et al (1997) to the star
formation history discussed by Madau (1997a) is not particularly good, but 
the observations are hardly completely determined at this point}. However,
there are some remaining puzzles. Foremost, in the hierarchical models, we
expect the growth of large disks to have occurred relatively recently 
(Efstathiou, this volume) and this should mirror the decline in the 
abundance of systems with irregular morphology (Baugh et al 1997). The 
presence of well-formed massive galaxies to at least $z$=0.8, with 
approximately the present comoving density (Lilly, this volume), suggests 
a more complex interpretation may be required. And, of course, many
suspect that the optically-detected trends may be only a lower limit to 
the star formation energy density that occurs over all wavelengths
(Rowan-Robinson, this volume).

\section{WHERE NEXT?} 

I would conclude that the much-heralded `synthesis' of theory and data
is premature. We have also only just scratched on the surface of the HDF 
data. Indeed, much of the recent progress is based merely on the 
{\it integrated} colours of HDF galaxies; we have hardly begun exploiting 
the 2-D resolved data which is the true benefit of using HST. I will
conclude by illustrating the next step in this sense which is based 
on preliminary work done in Cambridge (Abraham et al 1997, for a 
preliminary discussion see Abraham 1997).
 
Instead of using integrated 4-colour data to estimate redshifts, why
not use those sources for which spectroscopic redshifts are available
(Cowie et al 1996) and analyse the resolved pixel-by-pixel 4-colour data 
in constraining the star formation history of each galaxy and its
physical sub-components?  As most of the faint galaxies are extraordinarily
blue, the HDF colours are primarily sensitive to relative main sequence ages,
modulo small uncertainties in dust, the IMF and metallicity. These effects
would be worrisome if absolute ages were sought, but not if the goal is 
to examine the {\it relative spread} of burst ages ($\delta\,t/t$) across 
a galaxy in relation to its constituent spatial components. The technique 
can be tested for intermediate redshift spirals and ellipticals where 
sensible results emerge for the bulge and disk components and so, in Fig.~6, 
we illustrate how this method can be applied to the enigmatic high $z$
irregulars and so-called `chain galaxies'. Our resolved photometric analysis
determines that these systems are consistent with star formation progressing 
in distinct bursts occurring several hundred Myr apart, as might be expected 
if physically-independent young components are slowly assembling into 
larger structures.

\medskip
\centerline{\psfig{file=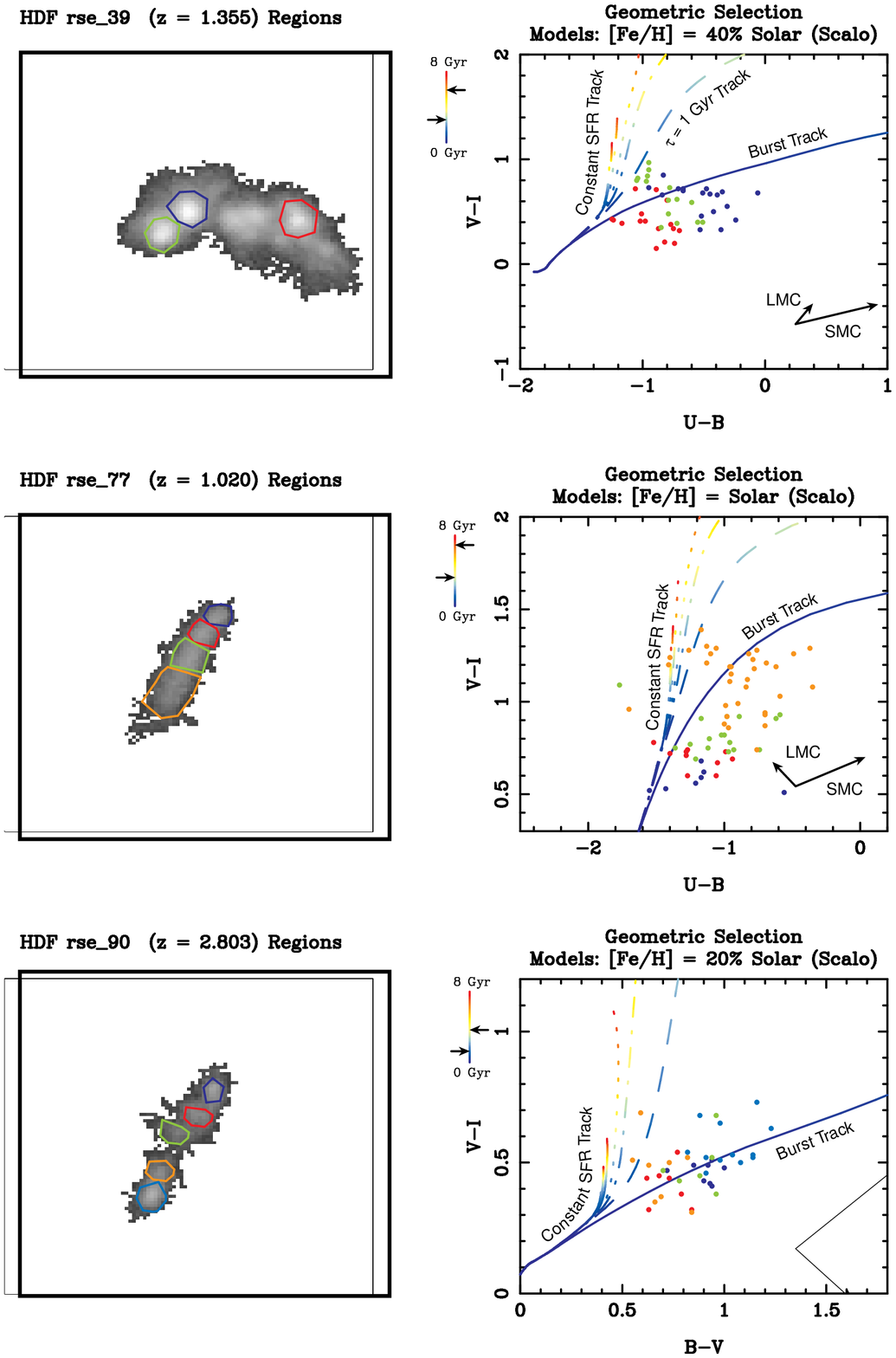,width=10cm}}
\noindent{{\it Figure 6: Resolved stellar populations for 3 high redshift
HDF galaxies from the forthcoming study of Abraham et al (for a preliminary 
discussion see Abraham (1997). The left panels show F814W images for 3
galaxies with Keck spectroscopy courtesy of Len Cowie. The right panels
show the pixel-by-pixel HDF colours grouped in physically-distinct
regions within each image as indicated by the colour scheme in the left
hand panels. Model tracks show evolutionary changes expected for various
stellar histories according to the cosmic clock in the vertical scale. 
The analysis suggest these three galaxies assembled from physically
and temporally-distinct stellar components.}
\medskip

Clearly this technique has limitations in terms of the history accessible when 
using optical data at high $z$, but the benefit of extending this kind of
study to include NICMOS $K$-band features sensitive to the mass of the
incoming components would be considerable, as would the possibility
of using GEMINI's integral field unit spectrographs to obtain the
associated dynamical data. Clearly this is a tough observational project
but I believe we should now complete the logical progression that
has delineated this subject observationally over the past two decades: 
from integrated photometry (counts) and spectroscopy (redshift surveys)
through to HST resolved imaging (MDS, HDF) to IFU-based spectroscopy. Such
exciting datasets we can fully expect to come shortly. They will demand 
equal progress in theoretical modelling which will have to become more 
realistic to give physical insight into the evolutionary processes which
are clearly occurring.

\section{CONCLUSIONS}

It's an exciting time to be doing cosmology! The last time we changed
government in England was when the first deep photographic counts were
published and Beatrice Tinsley suggested measuring the redshifts of a
sample of galaxies to $B$=21. She predicted a small fraction of the bluest
sources might be high redshift primordial galaxies. After 15 years of
ground-based work and only 4 years of post-refurbishment HST data, we
have clearly come a long way. The acceleration of this subject in the past
2 years owes a great deal to HST and, within that context, to the HDF itself.
There is much more data to come and much more physics to do. The much 
heralded `synthesis' of theory and data is premature in my view. So far 
we are mainly surveying. The more fundamental task of understanding will 
take considerably longer. When we finally get there, I believe we will 
all recall that moment when we first saw the spectacular image of the 
Hubble Deep Field.

\begin{acknowledgments}

I thank the organisers for inviting me to give this opening review
and for generous financial assistance. Full credit for the success of
the HDF must go to Bob Williams and his talented team at STScI.
\end{acknowledgments}

\end{document}